\documentclass[letterpaper,twocolumn,showpacs,prl,aps]{revtex4-1}

\usepackage{amsmath}  
\usepackage{amsfonts} 
\usepackage{graphicx} 
\usepackage{amssymb}
\usepackage{subfig}
\usepackage{float}
\usepackage{xcolor}

\begin{document}

\title{Negative group velocity and Kelvin-like wake pattern}

\author{Eugene B. Kolomeisky$^{1}$, Jonathan Colen$^{2}$, and Joseph P. Straley$^{3}$}

\affiliation
{$^{1}$Department of Physics, University of Virginia, P. O. Box 400714,
Charlottesville, Virginia 22904-4714, USA\\
$^{2}$Department of Physics, University of Chicago, 5720 South Ellis Ave, Chicago, Illinois 60637, USA\\
$^{3}$Department of Physics and Astronomy, University of Kentucky,
Lexington, Kentucky 40506-0055, USA}

\date{\today}

\begin{abstract}
Wake patterns due to a uniformly traveling source are a result of the resonant emission of the medium's collective excitations.  When there exists a frequency range where such excitations possess a negative group velocity, their interference leads to a wake pattern resembling the Kelvin ship wake:  while there are "transverse" and "divergent" wavefronts trailing the source, they are oriented oppositely to Kelvin's.  This is illustrated by an explicit calculation of "roton" wake patterns in superfluid $^{4}He$ where a Kelvin-like wake emerges when the speed of the source  marginally exceeds the Landau critical roton velocity.
\end{abstract}


\maketitle

Physics experiments typically measure a response of a system to an external disturbance.  A familiar type of disturbance is an object 
traveling uniformly relative to a medium.
This can give rise to a range of effects including Mach waves behind a supersonic projectile~\cite{LL6}, Cherenkov radiation emitted by a rapidly moving charge~\cite{LL8}, and ship waves~\cite{Kelvin, Lamb}, all of which are examples of coherent generation of the medium's collective excitations~\cite{wake_review}. 
 
If the disturbance is weak, the response is proportional to it and dictated by the properties of the medium.
This is the case for far-field wake patterns that are governed by the frequency spectrum $\Omega(\textbf{k})$ of the relevant collective excitations of the medium (here $\textbf{k}$ is the wave vector).  For any physical system, a wake is present whenever there is a wave mode whose phase velocity $\textbf{k}\Omega/k^{2}$ (here $k$=$|\textbf{k}|$) matches the projection of the velocity of the source $\textbf{v}$ onto the direction of radiation $\textbf{k}/k$ \cite{Lamb,wake_review}.  This implies that the wave pattern is stationary relative to the source.  For a source moving in the $+x$ direction with velocity $\textbf{v}$, this requires the existence of a wave vector $\mathbf{k}$ satisfying
\begin{equation}
\label{Cherenkov}
\Omega(\textbf{k})=\textbf{k}\cdot\textbf{v}\equiv kv\cos\varphi\equiv k_{x}v
\end{equation}     
where $v=|\textbf{v}|$ and $\varphi$ is the angle between the vectors $\textbf{k}$ and $\textbf{v}$.  Eq. (\ref{Cherenkov}) is the Mach-Cherenkov-Landau (MCL) resonant radiation condition, which also describes the 
onset of Landau damping in a plasma \cite{LL10} and the breakdown of superfluidity \cite{LL9}.

When the excitation spectrum is linear, 
\begin{equation}
\label{sound}
\Omega=uk
\end{equation}
where $u$ is the speed of sound (or light), the MCL condition (\ref{Cherenkov}) becomes $\cos \varphi=u/v$.  It can be satisfied only if $u\leqslant v$, i.e. if the source is supersonic (or superluminal).  

Contrarily, the spectrum of gravity waves is \cite{LL6,Lamb}
\begin{equation}
\label{spectrum}
\Omega^{2}=gk
\end{equation}
where $g$ is the free fall acceleration.   Now the MCL condition (\ref{Cherenkov}) becomes $\cos\varphi=\sqrt{g/kv^{2}}$ and can  always be satisfied for sufficiently large wave number $k$.  As a result, the wake appears for any velocity.

The complexity of wake patterns is governed by the spatial dispersion, i.e. by the difference between the group and the phase velocities.  Specifically, the simple cone geometry of Mach-Cherenkov wakes is due to the dispersionless spectrum (\ref{sound}) while the "feathered" appearance of ship wakes has its origin in the dispersive character of gravity waves (\ref{spectrum}).  

Here we show that when the group velocity is negative (i.e. it is opposite in direction to the phase velocity) in a range of the wave vectors satisfying Eq.(\ref{Cherenkov}), the corresponding excitations interfere and create a wake pattern that resembles the Kelvin ship wake.  This effect does not rely on the specific form of the dispersion law $\Omega(\textbf{k})$ as long as the latter contains region(s) with negative group velocity.  It is also different from a reversed Cherenkov effect in "left-handed" materials \cite{negative} that we do not consider.

Several physical systems possess excitations which in certain frequency ranges are characterized by negative group velocity.  A classic example is the optical branch of vibrations in crystals \cite{history}.   We will illustrate the impact of negative group velocity on wake patterns both generally for an arbitrary isotropic medium and explicitly using the example of superfluid $^{4}He$.  Past studies of wakes in a superfluid $^4He$~\cite{Rica,Ber} overlooked this effect.

For small wave numbers $k$ the collective (or elementary) excitations in Bose liquids correspond to hydrodynamic sound waves with a linear spectrum (\ref{sound}) \cite{LL9}.  In liquid $^{4}He$ the function $\Omega(\textbf{k}) \equiv \Omega(k)$ reaches a maximum after an initial increase, followed by a "roton" minimum at some $k_{0}$ \cite{LL9} (incidentally, this is also the case in a dipolar quantum gas \cite{dipole}). The excitations with wave numbers sandwiched between these extrema are characterized by a negative group velocity $d\Omega/dk\equiv\Omega'(k)<0$.  

In the vicinity of $k=k_{0}$ it is customary to expand the function $\Omega(k)$ in powers of $k-k_{0}$:
\begin{equation}
\label{roton}
\Omega=\frac{\Delta}{\hbar}+\frac{\hbar (k-k_{0})^{2}}{2\mu}
\end{equation}
where $\Delta$, $\mu$ and $k_{0}$ are empirically known parameters which depend on the pressure \cite{pressure}.   Hereafter for illustration purposes and without the loss of generality we employ their values extrapolated to zero pressure \cite{LL9}.  The parameters of the roton part of the spectrum (\ref{roton}) can be conveniently assembled into a dimensionless combination
\begin{equation}
\label{parameter}
a=\frac{\hbar^{2}k_{0}^{2}}{\mu \Delta}\approx 31.
\end{equation}  
Additionally, it is useful to introduce a velocity scale 
\begin{equation}
\label{reverse_velocity}
v_{0}=\frac{\Delta}{\hbar k_{0}}\approx 60~m/s,
\end{equation}
which is the slope of the straight line connecting the origin $\Omega(0) = 0$ to the roton minimum $\Omega(k_0) = \Delta / \hbar$.

The excitation spectrum of superfluid $^{4}He$ ends  at $\Omega=2\Delta/\hbar$, $k=k_{e}\approx 3\times 10^{8}~cm^{-1}$ \cite{LL9}.  Below we largely focus on the source velocities $v\leqslant 2\Delta/\hbar k_{e}\approx 76~m/s$ so that the end part of the spectrum does not contribute into the wake pattern.  

Hereafter it is convenient to measure the wave number and the source velocity in units of $k_{0}$ and $v_{0}$, respectively.   The roton part of the spectrum (\ref{roton}) will then be given by
\begin{equation}
\label{roton_dimensionless}
\Omega=1+\frac{1}{2}a(k-1)^{2}.
\end{equation}
The MCL requirement (\ref{Cherenkov}) is satisfied above a critical velocity $v_c$ such as $\Omega = k v$ and $\Omega'=v$.  Subjecting Eq.(\ref{roton_dimensionless}) to these conditions determines the threshold velocity to generate a wake which is also Landau's critical roton velocity to destroy superfluidity in $^{4}He$:  
\begin{equation}
\label{roton_velocity}
v_{c}=a\left (\sqrt{1+\frac{2}{a}}-1\right )\approx 0.98~(59~m/s,\text{original units})
\end{equation}
These conditions also define a critical wave number where the group and the phase velocities coincide, $\Omega'=\Omega/k$,
\begin{equation}
\label{critical_wave_number}
k_{c}=\sqrt{1+\frac{2}{a}}\approx 1.03.
\end{equation}  
Thanks to the very large parameter $a$ (\ref{parameter}) (which further increases with the pressure \cite{pressure}) the velocities $v_{0}$ (\ref{reverse_velocity}) and $v_{c}$ (\ref{roton_velocity}), and respective wave numbers $k_{0}$ and $k_{c}$, are extremely close to each other.  

Superfluidity can also be destroyed by vortex loop excitations above a velocity that depends on the specific conditions of the flow.   In the past this made it impossible to attain the Landau critical velocity (\ref{roton_velocity}) in practice.  Subsequent experiments with isotopically pure $^{4}He$ \cite{vortex} indicated that the critical velocity of vortex nucleation $u_{c}$ is significantly larger than the Landau value (\ref{roton_velocity}) (at a pressure of $12$ bars and larger).   There also is numerical indication \cite{Ber} relating $u_{c}$ to the speed of sound $u>v_{c}$ (\ref{sound}).  Hereafter vortex loop excitations are omitted since already at a moderate pressure their role is negligible. 

\begin{figure*}
\centering
\includegraphics[width=2.0\columnwidth,keepaspectratio]{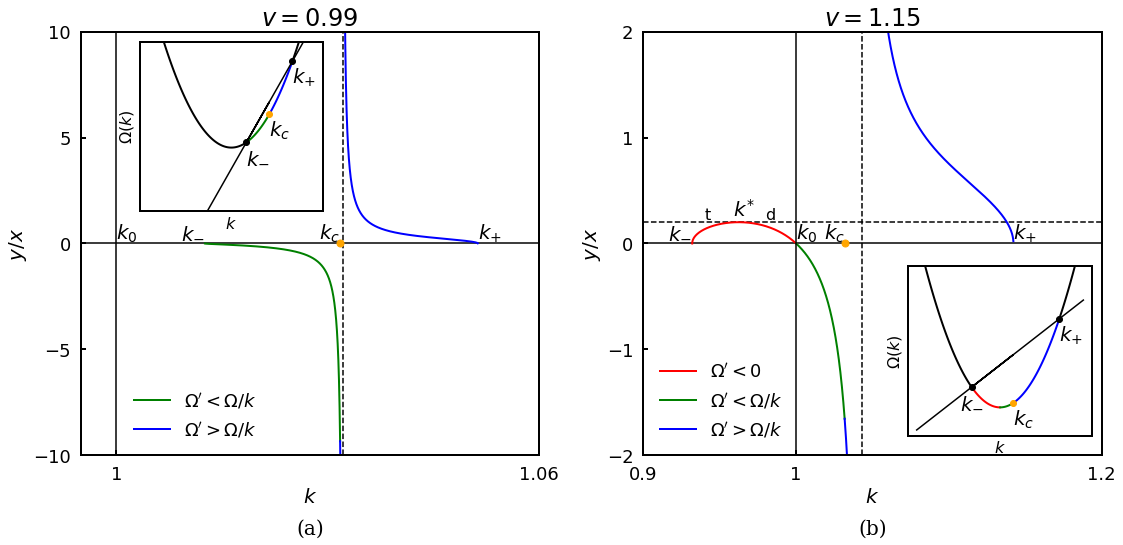}
\caption{The right-hand side of the equation of the stationary phase (\ref{stationary_phase_general}) in the roton approximation (\ref{roton_dimensionless}) with $v$ and $k$ measured in units of $v_0$ (\ref{reverse_velocity}) and $k_0$. (a) Source velocity $v=0.99$ ($v_{c} < v < 1$) 
versus (b) source velocity $v=1.15$ $(v > 1)$.  Insets illustrate the spectrum $\Omega(k)$ and the origin of the critical wave number $k_c$ (orange) (\ref{critical_wave_number}) and boundary values $k_{\pm}$ (\ref{end_values}) as solutions to $\Omega(k)=kv$. 
(b) includes waves with negative group velocity whose interference is responsible for transverse (t) and divergent (d) parts of the Kelvin-like wake in Figure \ref{wakes}b.}
\label{stphase}
\end{figure*}

When $v>v_{c}$, the MCL condition (\ref{Cherenkov}) holds within a range of the wave numbers $[k_{-},k_{+}]$.  In the roton approximation (\ref{roton_dimensionless}) the bounds $k_{\pm}$ are given by 
\begin{equation}
\label{end_values}
k_{\pm}=1+\frac{1}{a}\left ( v\pm \sqrt{v^{2}+2a(v-1)}\right )
\end{equation}
This shows the significance of the velocity $v_{0}$ (\ref{reverse_velocity}): if $v_{c}<v<v_{0}(=1)$, one has $k_{-}>k_{0}(=1)$ and all the waves with wave numbers in the $[k_{-},k_{+}]$ range have positive group velocity (inset to Figure \ref{stphase}a).   As the source velocity increases, the $[k_{-},k_{+}]$ interval widens and at $v=v_{0}(=1)$ one finds that $k_{-}=k_{0}(=1)$.  For $v>v_{0}$ the wave numbers in the $[k_{-},k_{0}]$ segment correspond to waves with negative group velocity (inset to Figure \ref{stphase}b).  

The resulting wake patterns can be understood through linear response theory \cite{LL5,Pines_Nozieres,KS,CK}.  We suppose that every particle of the medium is perturbed by a traveling external field of the potential energy $U(\textbf{r}-\textbf{v}t)$.  Then the average value of the Fourier transform of induced density due to the perturbation is given by $\delta \bar{n}(\omega,\textbf{k})=-2\pi\alpha(\textbf{k}\cdot \textbf{v},\textbf{k})U(\textbf{k})\delta(\omega-\textbf{k}\cdot\textbf{v})$ where $\alpha(\omega,\textbf{k})$ is a generalized susceptibility \cite{LL9,LL5, Pines_Nozieres} and $U(\textbf{k})$ is the Fourier transform of $U(\textbf{r})$.  Inverting the Fourier transform and changing the frame of reference to that of the source, $\textbf{r}-\textbf{v}t\rightarrow \textbf{r}$, the induced density is given by  
\begin{equation}
\label{wake_integral}
\delta \bar{n}(\textbf{r})=-\int\frac{d^{d}k}{(2\pi)^{d}}\alpha(\textbf{k}\cdot\textbf{v},\textbf{k})]U(\textbf{k})e^{i\textbf{k}\cdot\textbf{r}}
\end{equation}
where $d$ is the space dimensionality.  The susceptibility $\alpha(\omega,\textbf{k})$ has a pole at $\omega\equiv \textbf{k}\cdot\textbf{v}=\Omega(\textbf{k})$ \cite{LL9,Pines_Nozieres} which is the MCL condition (\ref{Cherenkov}).  

The wake pattern due to a point source ($U(\textbf{k})=const$) will be determined by a combination of Kelvin's stationary phase argument \cite{Kelvin,Lamb,wake_review} and numerical evaluation of the integral (\ref{wake_integral}) \cite{CK}, with
generalized susceptibility
\begin{equation}
\label{susceptibility}
\alpha(\omega,\textbf{k})\propto \frac{1}{(\omega+i0)^{2}-\Omega^{2}(\textbf{k})}.
\end{equation}

Far from the source, the phase $f=\mathbf{k}\cdot\mathbf{r}$ is large and the exponential in (\ref{wake_integral}) is highly oscillatory. This is where contributions of elementary plane waves interfere destructively leaving almost no net result, unless their wave vectors satisfy the MCL condition (\ref{Cherenkov}) and have a phase which is stationary with respect to $\mathbf{k}$.  This is the condition of constructive interference leading to a wake. 

We start by discussing two-dimensional wake patterns.  The wake is formed by interference of plane waves with positive $x$-components of the wave vector, $k_{x}>0$.  We also take $k_{y} > 0$ ($k_{y}<0$ contributions can be found via reflection $y\rightarrow -y$).  Then the wake is found at $y>0$ if the group velocity is positive or at $y<0$ if the group velocity is negative. The phase is given by
\begin{equation}
\label{phase_general}
f=\textbf{k}\cdot\textbf{r}=k_{x}x + k_{y}y = \frac{\Omega(k)}{v}x+\sqrt{k^{2}-\frac{\Omega^{2}(k)}{v^{2}}}y
\end{equation}
where we used the MCL condition (\ref{Cherenkov}) and $k^{2}=k_{x}^{2}+k_{y}^{2}$ to re-express $k_{x,y}$ in terms of $k$. Therefore the stationary phase condition $df/dk = 0$ becomes
\begin{equation}
\label{stationary_phase_general}
\frac{y}{x}=\frac{\Omega'\sqrt{v^{2}k^{2}-\Omega^{2}}}{\Omega\Omega'-v^{2}k}.
\end{equation}

Since the phase $f$ is constant along the wavefront, Eqs.(\ref{phase_general}) and (\ref{stationary_phase_general}) can be solved to give equations of the wavefronts in the parametric form:
\begin{equation}
\label{parametric_general}
x(k)=\frac{\Omega\Omega'-v^{2}k}{vk(\Omega'k-\Omega)}f,~~~~~ y(k)=\frac{\Omega'\sqrt{v^{2}k^{2}-\Omega^{2}}}{vk(\Omega'k-\Omega)}f.
\end{equation} 

Not only do Eqs.(\ref{phase_general})-(\ref{parametric_general}) describe wake patterns in two dimensions (\cite{Kelvin,Lamb,wake_review,KS,CK}), they also apply to the three-dimensional case.  The main difference is that the $xy$ plane becomes a plane intersecting the three-dimensional wake pattern along the path of the source.  Indeed, the different dimensionality does not change the phase function.    The way that the amplitude of the $d=3$ wake pattern falls off with distance will be different but the wave pattern itself will be the same, according to the stationary phase argument.  

The right-hand side (rhs) of Eq.(\ref{stationary_phase_general})
diverges at a wave number $k_{v}$ that can be bounded by evaluating $\Omega(k_{c})\Omega'(k_{c})-v^{2}k_{c}=k_{c}(v_{c}^{2}-v^{2})\leqslant0$ and $\Omega(k_{+})\Omega'(k_{+})-v^{2}k_{+}=k_{+}v[\Omega'(k_{+})-v]\geqslant 0$, meaning that $k_{c}\leqslant k_{v}\leqslant k_{+}$.
 
If the source velocity belongs to the $[v_{c},v_{0}]$ interval, the group velocity $\Omega'$ is positive in the $[k_{-},k_{+}]$ range. In the roton approximation (\ref{roton_dimensionless}) the behavior of the rhs of Eq.(\ref{stationary_phase_general}) is shown in Figure \ref{stphase}a.  Here $y/x$ is a monotonically decreasing function of $k$ both in the $[k_{-},k_{v}]$ range where $y/x<0$ and in the $[k_{v},k_{+}]$ interval where $y/x>0$.  Thus the wake is present both ahead ($x>0$) and behind ($x<0$) the source.  
\begin{figure*}
\centering
\includegraphics[width=2.0\columnwidth, keepaspectratio]{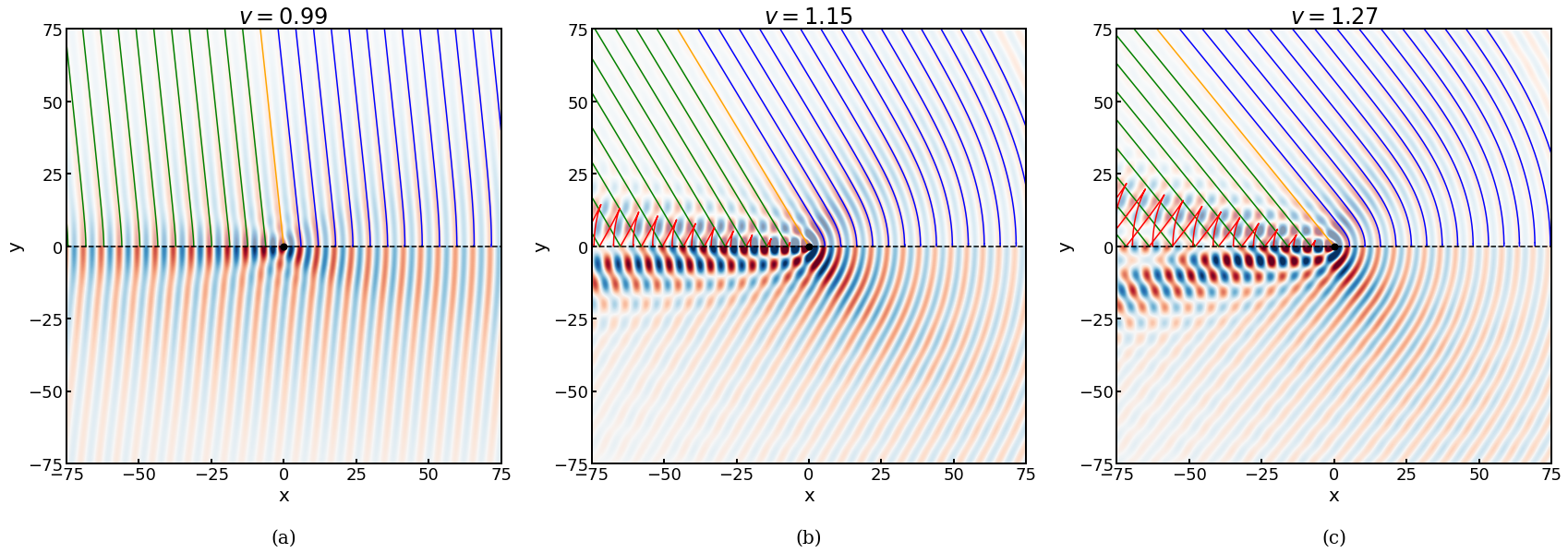}
\caption{Numerically evaluated two-dimensional roton wake patterns (\ref{wake_integral}) and (\ref{susceptibility}) due to a point source traveling to the right for different velocities measured in units of $v_0$ (\ref{reverse_velocity}). The unit of length is $1/k_{0}$,
and the color field indicates the size of the waves (arbitrary units).  The predictions of the stationary phase argument, Eqs.(\ref{parametric_general}), (\ref{front1}
-\ref{front2}), and (\ref{marginal}), color-matching Figure \ref{stphase}, are overlaid for $y>0$ where the wake pattern is made more transparent for visibility. 
The source velocities in parts (a) and (b) match Figures \ref{stphase}a and \ref{stphase}b.  Kelvin-like wakes, highlighted by a series of red triangles tracing wavecrests, are visible in parts (b) and (c) for $x<0$. }
\label{wakes}
\end{figure*}

Moreover, since for $k>k_{c}$ the group velocity $\Omega'$ is larger than the phase velocity $\Omega/k$, while for $k<k_{c}$, the opposite is true, $\Omega'<\Omega/k$, and signs of $y$ and $\Omega'$ coincide, the second of Eqs.(\ref{parametric_general}) implies that there are two families of wavefronts distinguished by the choice of the phase:
\begin{equation}
\label{front1}
f=2\pi n,~\text{if}~k\in[k_{c},k_{+}]~(\Omega'>\Omega/k)
\end{equation}
\begin{equation}
\label{front2}
f=-2\pi n,~\text{if}~~k\in[k_{-},k_{c}]~(\Omega'<\Omega/k)
\end{equation}  
where $n$ takes on positive integers.   Since for the first of these the rhs of the stationary phase condition (\ref{stationary_phase_general}) can be both positive and negative (blue-colored parts of the curves in Figure \ref{stphase}a), the wavefronts (\ref{parametric_general}) and (\ref{front1}) are found both at $x<0$ and $x>0$.  Specifically, $k=k_{v}$, the point of divergence of $y/x$, corresponds to $x=0$.  On the other hand, the wavefronts described by Eqs.(\ref{parametric_general}) and (\ref{front2}) are only found at $x<0$ because in the range of wave numbers (\ref{front2}) the rhs of Eq.(\ref{stationary_phase_general}) is negative (green-colored part of the curves in Figure \ref{stphase}a).

The spatial periodicities along the central line $y=0$ can be found by setting $k=k_{\pm}$ in the first of Eqs.(\ref{parametric_general}):
\begin{equation}
\label{periodicities}
\Delta x_{\pm}=\frac{2\pi}{k_{\pm}},
\end{equation}

The two wavefront families are separated by a marginal wavefront that can be obtained by evaluating Eq.(\ref{stationary_phase_general}) at $k=k_{c}$ (orange dot in Figure \ref{stphase}a):
\begin{equation}
\label{marginal}
\frac{y}{x}=-\frac{v_{c}}{\sqrt{v^{2}-v_{c}^{2}}}.
\end{equation}
This is a Mach-Cherenkov cone with the critical velocity $v_{c}$ playing a role of the limiting velocity.  The same equation describes the asymptotic behavior of the wavefronts (\ref{parametric_general}), (\ref{front1}) and (\ref{front2}) far away from the source.  
 
In the roton approximation (\ref{roton_dimensionless}) the resulting wake pattern is shown in Figure \ref{wakes}a where predictions of the stationary phase analysis, Eqs.(\ref{parametric_general}), (\ref{front1}
-\ref{front2}), and (\ref{marginal}) (color matching Figure \ref{stphase}a) are overlaid (for $y>0$) with results of numerical evaluation of the integral (\ref{wake_integral}) for $d=2$.  The latter additionally supplies the information about the size of the waves (in arbitrary units) varying from crests (deep red) to troughs (deep blue).  

A wake pattern of this kind will be found in any system that has an analog of the Landau velocity if the source velocity only slightly exceeds $v_{c}$.  The pattern can be deduced via an expansion of a dispersion law $\Omega(k)$ about $k=k_{c}$ up to the second order in $(k-k_{c})$.  Such a wake is indeed observed in water where due to the effects of capillarity omitted in Eq.(\ref{spectrum}), one finds $v_{c}\approx 23~cm/s$ \cite{Lamb}.
    
As $v\rightarrow v_{c}+0$, the wake in Figure \ref{wakes}a transforms into a one-dimensional periodic pattern of induced density $\delta \overline{n}(x)$ with the period $\Delta x=2\pi /k_{c}$, which was also predicted to occur when the velocity of a bulk uniform flow slightly exceeds the Landau critical roton velocity \cite{Pitaevskii}.
 
If $v>v_{0}$, the conclusions found for the $v_{c}<v<v_{0}$ regime carry over for waves in the range $[k_{0}, k_{+}]$ where the group velocity $\Omega'$ remains positive. This is illustrated in Figure \ref{stphase}b where we plotted Eq.(\ref{stationary_phase_general}) for $v>v_{0}$ in the roton approximation (\ref{roton_dimensionless}).  The behavior of $y/x$ in the $[k_{0},k_{+}]$ range is qualitatively the same as that in Figure \ref{stphase}a.  This explains a part of the wake pattern in Figure \ref{wakes}b that is qualitatively similar to that in Figure \ref{wakes}a.  

For waves in the range $[k_{-}, k_0]$, the group velocity $\Omega'$ is negative, and the rhs of Eq.(\ref{stationary_phase_general}) is positive.  Since the signs of $\Omega'$ and $y$ coincide, part of the wake described by Eqs.(\ref{parametric_general}) and (\ref{front2}) due to excitations of the $[k_{-},k_{0}]$ range will still be found behind the source, $x<0$, but at $y<0$.  Moreover, a function $y/x$ positive in the $[k_{-},k_{0}]$ interval and vanishing at its ends must have at least one maximum within it.  This general reasoning is illustrated in Figure \ref{stphase}b which shows that in the roton approximation the rhs of Eq.(\ref{stationary_phase_general}) is positive in the $[k_{-},k_{0}]$ range (red curve) and has a maximum at a wavenumber $k^{*}$.  

If $0\leqslant y/x<(y/x)^{*}\equiv(y/x)(k^{*})$ (shown in Figure \ref{stphase}b as a dashed horizontal line)  Eq.(\ref{stationary_phase_general}) has three solutions.  One of them immediately to the left of $k=k_{+}$ corresponds to already discussed wavefronts (\ref{parametric_general}) and (\ref{front1}).  The remaining two solutions within the $[k_{-},k_{0}]$ interval, one to the left and one to the right of $k=k^{*}$, are new.  As $y/x$ increases approaching $(y/x)^{*}$ the two solutions tend to each other, join at $y/x=(y/x)^{*}$, and none are found if $y/x>(y/x)^{*}$.  This part of the wake pattern confined within an angle $2\arctan(y/x)^{*}$ is represented in Figures \ref{wakes}b and \ref{wakes}c by a series of curved red triangles.  

There is a substantial similarity between this pattern and the classic Kelvin ship wake formed behind a point pressure source uniformly traveling a calm water surface \cite{Kelvin,Lamb}.  Specifically, Kelvin's terminology of "transverse" and "divergent" \cite{Kelvin} wavefronts applies:  interfering excitations with wave numbers in the $[k_{-},k^{*}]$ ($[k^{*},k_{0}]$) range produce the transverse (divergent) wavefronts which smoothly (discontinuosly) connect the edges of the pattern $y/x=\pm(y/x)^{*}$ across the central line $y=0$.  The notable difference from Kelvin's is that the pattern is reversed (i.e. it faces \textit{away} from the source) as the wake is formed by elementary waves with negative group velocity.  

Like the Landau critical roton velocity $v=v_{c}$, the instant $v=v_{0}$ where elementary waves having negative group velocity start participating in forming wake pattern, represents a critical phenomenon.  It is expected that it will be accompanied by an increase in the wave resistance which we are planning to study in the future.   

As the source velocity increases beyond $v=v_{0}$, the size of the waves increases, the opening angle of the Kelvin-like wake grows and the roton approximation (\ref{roton}) eventually breaks down.  However qualitatively the wake pattern will not change as long as $v\lesssim 76~m/s$; the pattern in Figure \ref{wakes}c corresponds to $v=76~m/s$.  At larger velocities the wake pattern will acquire new elements due to contributions coming from the end part of the spectrum.  At the same time the Kelvin-like feature will persist and widen with $v$ until the source velocity exceeds about $167~m/s$.  This is when the lower bound of the MCL interval $k_{-}$ coincides with location of the maximum of $\Omega(k)$ and the range of wave numbers with $\Omega'<0$ is largest.    

To summarize, we demonstrated that if in response to a small heavy uniformly traveling source a medium radiates excitations possessing negative group velocity,  their interference leads to a reversed Kelvin-like wake.  In superfluid $^{4}He$ this effect should be commonplace because the Landau critical roton velocity of $v_{c}=59~m/s$ is extremely close to the Kelvin threshold of $v_{0}=60~m/s$.  At the same time the details of the Kelvin feature may be somewhat obscured due to interference with overlapping "green" family of wavefronts, Figures \ref{wakes}b and \ref{wakes}c. A Kelvin-like wedge should be observable by light scattering techniques while the fine structure of all the discussed wake patterns can be only resolved with the help of $X$-ray or neutron scattering.

This work was performed in part at Aspen Center for Physics, which is supported by National Science Foundation grant PHY-1607611.


\begin{thebibliography}{15}

\bibitem{LL6} L. D. Landau and E. M. Lifshitz, \textit{Fluid Mechanics:  Volume 6} (Course of Theoretical
Physics) (Pergamon, Oxford, 1987), Sections 12 and 82.

\bibitem{LL8} L. D. Landau and E. M. Lifshitz, \textit{Electrodynamics of Continuous Media: Volume 8} (Course of Theoretical
Physics) (Pergamon, Oxford, 1984), Section 115.

\bibitem{Kelvin}  L. Kelvin, \textit{On Ship Waves}, Proc. Inst. Mech. Eng. \textbf{3}, 409 (1887).

\bibitem{Lamb} H. Lamb, \textit{Hydrodynamics} (6th ed., Cambridge University Press, 1975), Chapter IX.

\bibitem{wake_review} I. Carusotto and G. Rousseaux, in \textit{Analogue Gravity Phenomenology}, D. Faccio \textit{et al.} (eds), Lecture Notes in Physics, Chapter 6, p.109.

\bibitem{LL10}  E. M. Lifshitz and L. P. Pitaevskii, \textit{Physical Kinetics:  Volume 10} (Course of Theoretical
Physics) (Butterworth-Heinemann, 1981), Section 30.

\bibitem{LL9}  E. M. Lifshitz and L. P. Pitaevskii, \textit{Statistical Physics, Part 2:  Volume 9} (Course of Theoretical Physics) (Butterworth-Heinemann, 1980), Chapters III and IX.

\bibitem{negative}  V. E. Pafomov, \textit{Transition radiation and Cerenkov radiation}, Zh. Eksp. Teor. Fiz. \textbf{36}, 1853 (1959) [Sov. Phys. JETP \textbf{9}, 1321 (1959)];  V. G. Veselago, \textit{The electrodynamics of substances with simultaneously negative values of $\epsilon$ and $\mu$}, Usp. Fiz. Nauk \textbf{92}, 517 (1967) [Sov. Phys. Usp. \textbf{10}, 509 (1968)];  \textit{Waves in metamaterials: their role in modern physics}, Usp. Fiz. Nauk \textbf{181}, 1201 (2011) [Phys.-Usp. \textbf{54}, 1161 (2011)], and references therein.

\bibitem{history} A brief history of negative group velocity is outlined in K. T. McDonald, \textit{Negative group velocity}, Am. J. Phys. \textbf{69}, 607 (2001).

\bibitem{Rica}  Y. Pomeau and S. Rica, \textit{Model of Superflow with Rotons}, Phys. Rev. Lett. \textbf{71}, 247 (1993).

\bibitem{Ber}  N. G. Berloff and P. H. Roberts, \textit{Roton creation and vortex nucleation in superfluids}, Phys. Lett. A \textbf{274}, 69 (2000).

\bibitem{dipole}  L. Chomaz, R. M. W. van Bijnen, D. Petter, G. Faraoni, S. Baier, J. H. Becher, M. J. Mark, F. W\"achtler, L. Santos and F. Ferlaino, \textit{Observation of roton mode population in a dipolar quantum gas}, Nature Physics \textbf{14}, 442 (2018), and references therein.


\bibitem{pressure}  H. Godfrin, K. Beauvois, A. Sultan, E. Krotscheck, J. Dawidowski, B. F\aa k, and J. Ollivier, \textit{Dispersion relation of Landau elementary excitations and thermodynamic properties of superfluid
$^{4}He$}, Phys. Rev. B \textbf{103}, 104516 (2021).

\bibitem{vortex}  R. M. Bowley, P. V. E. McClintock, F. E. Moss, and P. C. E. Stamp, \textit{Vortex Nucleation in Isotopically Pure Superfluid $^{4}He$}, Phys. Rev. Lett.  \textbf{44}, 161 (1980); P. C. Hendry, N. S. Lawson, P. V. E. McClintock, C. D. H. Williams, and R. M. Bowley, \textit{Macroscopic Quantum Tunneling of Vortices in He II}, Phys. Rev. Lett. \textbf{60}, 604 (1988).  

\bibitem{LL5}  L. D. Landau and E. M. Lifshitz, \textit{Statistical Physics, Part 1: Volume 5} (Course of Theoretical
Physics) (Butterworth-Heinemann, 1980), Section 123.

\bibitem{Pines_Nozieres}  D.Pines and P. Nozi\`eres, \textit{The Theory of Quantum Liquids}, Vol.1 (Benjamin, New York 1966), Chapter 2.

\bibitem{KS}  E. B. Kolomeisky and J. P. Straley, \textit{Kelvin-Mach Wake in a Two-Dimensional Fermi Sea}, Phys. Rev. Lett. \textbf{120}, 226801 (2018).

\bibitem{CK}  J. Colen and E. B. Kolomeisky, \textit{Kelvin-Froude wake patterns of a traveling pressure disturbance}, Eur. J. Mech. /B Fluids \textbf{85}, 400 (2021).

\bibitem{Pitaevskii}  L. P. Pitaevski\u{i}, \textit{Layered structure of superfluid $^{4}He$ with supercritical motion}, JETP Lett. \textbf{39}, 511 (1984) [Pis'ma Zh. Eksp. Teor. Fiz. \textbf{39}, 423 (1984)];  F. Ancilotto, F. Dalfovo, L. P. Pitaevskii, and F. Toigo, \textit{Density pattern in supercritical flow of liquid $^{4}He$}, Phys. Rev. B \textbf{71}, 104530 (2005).


\end{thebibliography}
\end{document}